\documentclass[12pt,preprint2]{aastex}

\newcommand{\chandra}{{\em Chandra\ }}
\newcommand{\chandrano}{{\em Chandra}}
\newcommand{\rxte}{{\em RXTE\ }}
\newcommand{\rxteno}{{\em RXTE}}
\newcommand{\rosat}{{\em ROSAT\ }}
\newcommand{\rosatno}{{\em ROSAT}}
\newcommand{\BeppoSAX}{{\em BeppoSAX\ }}
\newcommand{\BeppoSAXno}{{\em BeppoSAX}}
\newcommand{\ergcms}{erg~cm$^{-2}$~s$^{-1}$}
\newcommand{\ergss}{erg~s$^{-1}$}
\newcommand{\cmsq}{cm$^{-2}$}
\newcommand{\cxo}{CXOU\ J205847.5+414637}
\begin{document}
\title{Discovery of a Be/X-ray Binary Consistent with the Position of GRO
J2058+42}
\author{Colleen A. Wilson\altaffilmark{1}, Martin C. Weisskopf\altaffilmark{1},
Mark H. Finger\altaffilmark{2}}
\affil{XD 12 Space Science Branch, National Space Science and 
Technology Center, 320 Sparkman Drive, Huntsville, AL 35805}
\email{colleen.wilson@nasa.gov}
\author{M.J. Coe}
\affil{School of Physics and Astronomy, The University, Southampton, SO17 1BJ,
England}
\author{Jochen Greiner}
\affil{Max-Planck-Institut für Extraterrestrische Physik, Giessenbachstrasse, 
85748 Garching, Germany}
\author{Pablo Reig, Giannis Papamastorakis}
\affil{Physics Department, University of Crete, 71003, Heraklion, Greece and 
IESL, Foundation for Research and Technology, 71110, Heraklion, Greece}
\altaffiltext{1}{NASA's Marshall Space Flight Center}
\altaffiltext{2}{Universities Space Research Association}
\begin{abstract}
GRO J2058+42 is a 195 s transient X-ray pulsar discovered in 1995 with BATSE. In
1996, \rxte located GRO J2058+42 to a 90\% confidence error circle with a
$4\arcmin$\ radius. On 20 February 2004, the region including the error circle
was observed with \chandra ACIS-I. No X-ray sources were detected within the error 
circle, however, 2 faint sources were detected in the ACIS-I field-of-view. We 
obtained optical observations of the brightest object, \cxo, 
that had about 64 X-ray counts and was just $0.3\arcmin$\ outside the error 
circle. The optical spectrum contained a strong H$\alpha$ line and corresponds 
to an infrared object in the 2MASS catalog, indicating a Be/X-ray binary system. 
Pulsations were not detected in the \chandra observations, but similar flux
variations and distance estimates suggest that \cxo\ and GRO
J2058+42 are the same object. We present results from the Chandra 
observation, optical observations, new and previously unreported \rxte
observations, and a reanalysis of a \rosat observation.
\end{abstract}

\section{Introduction}
\subsection{Be/X-ray binaries}
The majority of the known accretion-powered pulsars are 
transients in binary systems with Be (or Oe) stars \citep{Apparao94,Coe00}. 
Be stars are main-sequence B stars 
showing emission in the Balmer lines \citep{Porter03, Slettebak88}. 
This line emission, and a strong infrared excess in comparison to normal stars
of the same stellar type, is associated with circumstellar material which is
being shed by the star into its equatorial plane. The exact nature of the mass
loss process is unknown, but it is thought to be related to the rapid rotation,
which is typically near or above 70\% of the critical break-up velocity
\citep{Porter96, Townsend04}. Near the Be star the equatorial material probably 
forms a quasi-Keplerian disk \citep[e.g.,][]{Quirrenbach97, Hanuschik96},
believed to fuel the X-ray outbursts.

In the ``normal" outbursting behavior (type I) of a Be/X-ray binary,  
series of moderate luminosity ($10^{35-37}$ \ergss) 
outbursts occur, with each outburst near the periastron passage phase
of the system's wide eccentric orbit \citep{Stella86, Bildsten97}. 
Some systems also show infrequent ``giant" (type II) outbursts, with 
luminosities of $10^{38}$ \ergss\ or greater. 
In  most systems there is no clear correlation between X-ray outbursts and 
optical activity within single outbursts.
The X-ray activity however, follows the long-term optical activity cycle 
of the Be star, in the sense that no outbursts occur in periods where optical 
indicators of the Be star disk, such as $H\alpha$ emission, have disappeared.
Periods of X-ray quiescence when the Be disk is present are also observed;
these may be due to the truncation of the Be star disk well 
within the neutron star orbit \citep{Negueruela01a, Negueruela01b}. 

\subsection{GRO J2058+42}
GRO J2058+42 was discovered as a 198 s pulsar with the Burst and Transient
Source Experiment (BATSE) on the {\em Compton Gamma Ray Observatory} in 1995, 
during a giant outburst \citep{Wilson95,Wilson98}. In the giant outburst, the 
pulsar spun-up from a period of 198 s to a period of about 196 s. The giant
outburst was followed by a series of normal outbursts, spaced at 55-day
intervals. For the first 11 normal outbursts, there was an alternating pattern 
in 20-50 keV pulsed flux (i.e. odd outbursts, counting from the giant outburst,
were brighter than even outbursts). However, the All-Sky Monitor (ASM) on the 
{\em Rossi X-ray Timing Explorer (RXTE)}, which observed the phase averaged flux
in the 2-10 keV band for 10 of these outbursts, did not see such a pattern, 
suggesting that the pattern may have been caused by changes in the spectrum or 
pulsed fraction.  Later, the outbursts continued, but the odd-even pattern 
stopped in the 20-50 keV pulsed flux. All of the outbursts showed frequency 
increases which, if attributed to mainly orbital effects, required 
an orbital period of 55 days \citep{Wilson98, Wilson00}. The outburst behavior,
specifically the giant outburst followed by a series of normal outbursts, 
suggested a Be star companion. 

Initially, GRO J2058+42 was approximately located with BATSE \citep{Wilson95} 
and with the Oriented Scintillation Spectroscopy Experiment \citep{Grove95}. In
1996 November, scanning observations with the \rxte Proportional Counter Array 
(PCA) further reduced the error region to a 90\% confidence 4\arcmin\ radius 
error circle centered on $\alpha = 314.75\arcdeg$, $\delta = 41.72\arcdeg$ 
\citep{Wilson96}. \citet{Castro96} reported two optical objects, with magnitudes
18.5 and 19.5, respectively, in the \rxte 
error circle at $\alpha = 314.7417\arcdeg$, $\delta = +41.7183\arcdeg$ and 
$\alpha = 314.7821\arcdeg$, $\delta = +41.7461\arcdeg$\ (J2000, uncertainty 
$\pm 1\arcsec$) on October 15, 1996. However, we found no X-ray object at 
either position with \rosat or \chandra. The \rosat HRI catalog, available 
through HEASARC
(ROSATHRITOTAL)\footnote{\url{http://heasarc.gsfc.nasa.gov/W3Browse/rosat/\\roshritotal.html}}
and through MPE (1RXH catalog)\footnote{\url{http://www.xray.mpe.mpg.de/cgi-bin/rosat/src-browser}}, 
lists 7 sources in the \rosat HRI field of view from a 1997 June 23 observation
centered on the \rxte position, but only 3 of these sources were unique and none
were within the \rxte error circle. Recently, \citet{Reig04} performed an 
optical photometric and spectroscopic analysis of the field around GRO J2058+42
from the Skinakas observatory and suggested the star approximately located at 
$\alpha = 314.697\arcdeg$, $\delta = +41.777\arcdeg$ (J2000) as its optical 
counterpart. 

In this paper we report on Chandra observations, archival
and new \rxte observations, archival \rosat observations, and give 
the details of the optical observations and analysis. The combination of 
archival and new X-ray and optical data allowed us to pin down the X-ray source
and identify its likely optical counterpart with a moderately reddened Be star.
We discuss the possible association between a \chandra source and GRO
J2058+42. 

\section{GRO J2058+42 X-ray Observations with \rxte}
\begin{figure}[!h]
\plotone{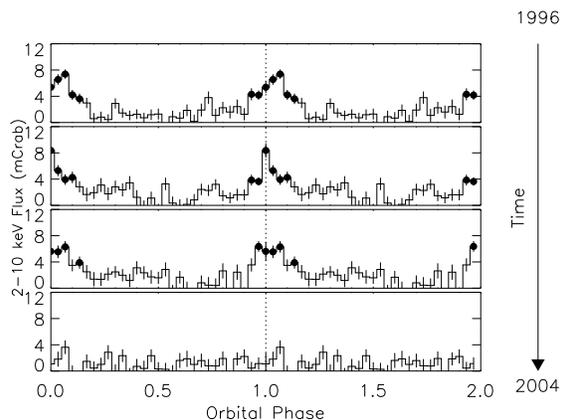}
\caption{\rxte ASM data for GRO J2058+42 divided in 4 equal $\sim 2.1$ year intervals (1996
February-1998 March, 1998 March-2000 May, 2000 May-2002 June, and 2002 June-2004
July) and folded at the 55.03 day period reported in \citet{Wilson00}. Filled 
circles denote $\gtrsim 3 \sigma$ detections.\label{fig:asm}}
\end{figure}

Figure~\ref{fig:asm} shows all-sky monitor (ASM) data from the entire \rxte 
mission through 2004 July divided into 4 equal $\sim 2.1$ year intervals and 
epoch-folded using the empheris $T ={\rm MJD}\ 50411.3+55.03 N$ \citep{Wilson00}.
From 1996-2002, GRO J2058+42 was detectable with the ASM from about 3-5 days 
before until about 6-8 days after the predicted outburst peak.

In early 1998, two outbursts of GRO J2058+42 were observed with the \rxte PCA
and High Energy Timing Experiment (HEXTE). Fourteen observations were obtained 
from 1998 January 23 - February 24 and 13 observations from 1998 March 18-30. 
Details including start and stop times for each observation can be found through the 
HEASARC\footnote{\url{http://heasarc.gsfc.nasa.gov}}. These observations spanned
from about 16 days until 2 days before the outburst peak. Pulsations were 
detected from about 6.5 days before the peak until the observations ceased
with r.m.s. pulsed fluxes of 0.2-1.4 mcrab over the full 2-60 keV PCA energy 
range. 

We analyzed energy spectra from archival \rxte data for these two outbursts to
look for spectral variations with intensity. Previous efforts 
\citep{WilsonHodge99,Wilson00} reported only on spectral analysis of the 
brightest of these observations. We analyzed archival PCA Standard2 data and 
HEXTE science event mode E\_8us\_256\_DX1F data for all 27 of the observations,
using FTOOLs\footnote{\url{http://heasarc.gsfc.nasa.gov/ftools}} version 5.3 
\citep{Blackburn95} to generate spectra and response files. Background spectra 
for the PCA observations were generated using the ``faint" source models. In 
XSPEC\footnote{\url{http://heasarc.gsfc.nasa.gov/docs/xanadu/xspec/\\index.html}}
version 11.3.1 \citep{Arnaud96}, the fits included PCA data from 2.7-25 keV and
HEXTE data from 11-50 keV. All of the observations were well fitted with an 
absorbed thermal Bremsstrahlung model (phabs\footnote{Abundances are given by
\citet{Anders89} and cross-sections by \citet{Bal92} with a new He cross-section
based on \citet{Yan98} in the phabs model, used in fits to \rxte and \chandra
data. Thompson scattering is not included in the phabs model.}*bremss) with reduced $\chi^2$ values
in the range 0.8-1.3 for 126 degrees of freedom. More sophisticated models, such
as a power-law with a high-energy cutoff, were not used because the 
Bremsstrahlung model adequately fit the data and provided a simple measure of 
spectral hardness across all source intensities. The data were better fitted 
with the Bremsstrahlung model than with a power-law at higher intensities. 
Estimated 2-50 keV fluxes ranged from $6.8 \times 10^{-13}$ \ergcms\ to 
$2.6 \times 10^{-10}$ \ergcms. From about 8-9 days before the predicted peak, 
the flux began to increase, rising above a few $\times 10^{-12}$ \ergcms. The 
spectra were well determined for fluxes above about $2 \times 10^{-11}$ \ergcms,
from about 6.5 days before the predicted peak, when pulsations were also 
detected. The absorption, $N_{\rm H}$, appeared to be constant with best fit 
values in the range $(4.6-5.4) \times 10^{22}$ \cmsq. The temperature, $kT$, 
increased as the intensity increased, with best fit values from $10.3 \pm 0.5$ 
keV at $2.9 \times 10^{-11}$ \ergcms\ to $22.2 \pm 0.4$ keV at $2.6 \times 
10^{-10}$ \ergcms.  

Because of the 1\arcdeg\ FWHM field-of-view of the PCA, these spectra included 
emission from both GRO J2058+42 and the Galactic ridge. The diffuse emission 
from the Galactic ridge is not well described at high Galactic longitudes (for 
GRO  J2058+42 $\ell = 83.6\arcdeg$). In fact, the models of \citet{Valinia98} 
and \citet{Rev03} predict a Galactic ridge flux of $\sim 2 \times 10^{-11}$ 
\ergcms\ (2-10 keV), taking into account GRO J2058+42's galactic latitude ($b =
-2.6\arcdeg$), an order of magnitude higher than our faintest flux measurements for
the GRO J2058+42 region. The column density measured with \rxte was nearly an 
order of magnitude larger than the Galactic value of $7 \times 10^{21}$ \cmsq\ 
calculated with Colden, the Galactic Neutral Hydrogen Density 
Calculator\footnote{\url{http://asc.harvard.edu/toolkit/colden.jsp}}, based on 
\citet{Dickey90}, suggesting there was a significant column of material
intrinsic to the GRO J2058+42 system.  

In preparation for our \chandra observations, \rxte observed the GRO J2058+42 
region in a series of 27 observations from 2003 December 18 - 2004 January 1, 
spanning from 6.5 days before until 7.5 days after the predicted peak and in a 
second series of five very short ($\lesssim 1$ ks) observations from 2004 
February 15-19, from 2.5 days before until 1.5 days after the predicted peak. 
Pulsations were not detected in any of the 2003-2004 observations at or above 
the previous minimum detection of an r.m.s. pulsed flux of 0.2 mcrab in the full
2-60 keV band. Most of the observations in 2003 December were short, 
$\lesssim 3$ ks. For the eight longer observations from 2003 December, with 
exposure times of 6-14 ks, we fitted 2.7-15 keV PCA data in XSPEC with an 
absorbed thermal Bremsstrahlung model, with $N_{H}$ fixed at $2.1 \times 
10^{22}$ \cmsq, the best-fit value measured with \chandra (See 
Section~\ref{ss:xray_color_color}). Best fit temperatures ranged from 3-5 keV 
and 2-10 keV fluxes from $(2-3) \times 10^{-12}$ \ergcms, with no evidence of
a correlation. These fluxes should be treated as upper limits for the GRO 
J2058+42 flux, since no pulsations were detected with \rxte and the spectra also
include Galactic ridge emission. 

\section{X-ray Observations of\\ CXO J205847.51+4146373}
\subsection{\chandra}
On 20 February, 2004, \chandra observed the field containing GRO J2058+42 for
9.9 ks with the Advanced CCD Imaging Spectrometer (ACIS) array and the High 
Energy Transmission Grating (HETG) in the faint, timed-exposure mode with a 
frame time of 3.241 seconds. The Imaging Array (ACIS-I) field-of-view was 
centered on $\alpha = 314.75\arcdeg$, $\delta = +41.72\arcdeg$ (J2000), the 
center of the \rxte error circle for GRO J2058+42. The ACIS-I chips were on 
(I0, I1, I2, I3) and chips S2 and S3 were also active. Standard processing 
v.7.7.1 applied aspect corrections and compensated for spacecraft dither. 
Level 2 events were used in our analysis. Data in the energy range 0.5-8 keV 
were used for all analyses to reduce background. We simultaneously processed 
these data using Chandra Interactive Analysis of Observations 
(CIAO)\footnote{\url{http://cxc.harvard.edu/ciao}} version 3.1 and 
CALDB\footnote{See \url{http://asc.harvard.edu/CIAO/} for more information}
version 2.27 and with the techniques described in \citet{Swartz03}.

\subsubsection{Source Detection\label{ss:detect}}

\begin{deluxetable}{lccccccccc}
\tabletypesize{\scriptsize}
   \tablewidth{0pc}
   \tablecaption{Sources Detected with \chandra \label{source_table}}
    \tablehead{\colhead{NAME} & \colhead{R.A.} & \colhead{Dec} & 
    \colhead{$r_1^a$} & \colhead{$N^b$} & \colhead{$S/N^c$} & 
    \colhead{$r_2^d$} & \colhead{USNO$^e$} & \colhead{2MASS$^e$} & 
    \colhead{\rosat$^e$}    \\ 
     & \colhead{J2000} & \colhead{J2000} & \colhead{($\arcsec$)} & 
    & & \colhead{($\arcsec$)} & \colhead{($\arcsec$)} & \colhead{($\arcsec$)} 
     & \colhead{($\arcsec$)}}
   \startdata
  I01 & 314.67230\arcdeg & 41.58780\arcdeg & 10.6 & 10 & 2.9 & 4.2   &       &      &  \\
  \cxo\ & 314.69794\arcdeg & 41.77704\arcdeg & 3.7  & 64 & 7.3 & 1.4  & 0.37  & 0.35 & 0.48 \\
\enddata
\tablenotetext{a}{Extraction radius, defined as $2.5 \sigma$, where
$\sigma$ is defined in Section~\ref{ss:detect}.} 
\tablenotetext{b}{Approximate number of source counts.}
\tablenotetext{c}{Detection signal-to-noise ratio.}
\tablenotetext{d}{X-ray position uncertainty (99\% confidence radius).}
\tablenotetext{e}{Radial separation between X-ray position and cataloged 
position of counterpart.}
\end{deluxetable}

We used the same source finding techniques as described in \citet{Swartz03}
with the circular-Gaussian approximation to the point spread function, and a
minimum signal-to-noise (S/N) ratio of 2.6 resulting in much less than 1
accidental detection in the field. The corresponding background-subtracted 
point source detection limit is $\sim$10 counts which corresponds to a flux 
(0.5-8.0 keV) of about $7\times 10^{-15}$ \ergcms\ for an unabsorbed power-law of
spectral index $-1.5$. Two sources were detected. Table~\ref{source_table} gives 
the source positions, the extraction radius, the net counts, the 
signal-to-noise ratio, and the associated uncertainty in the X-ray position. 
Table~\ref{source_table} also indicates whether or not the source is 
identified with a counterpart in either the United States Naval Observatory 
Catalog \citep[USNO-B1.0,][]{Monet03}, the Two Micron All Sky Survey (2MASS) 
and the various \rosat catalogs and lists the angular separation between
the X-ray position and the optical/infrared/\rosat positions.

The 99\% confidence positional uncertainty listed in column 7 of 
Table~\ref{source_table} is given by $r=3.03(\sigma^2/N + \sigma^2_o + 
\sigma^2_1)^{1/2}$ where $\sigma$ is the standard deviation of the circular 
Gaussian that approximately matches the PSF at the source location, $N$ is the 
aperture-corrected number of source counts for a 2.5 $\sigma$ extraction radius,
$\sigma_o$ represents the uncertainty in the image reconstruction
produced by irremovable contributions to the relative aspect solution, and
$\sigma_1$ accounts for the uncertainties in the absolute aspect solution.
These latter two uncertainties are discussed in \cite{Weisskopf03}\footnote{See
also \url{http://cxc.harvard.edu/cal/ASPECT/celmon/} and 
\url{http://cxc.harvard.edu/cal/ASPECT/img\_recon/\\report.html}} .
We have used 0.2\arcsec as a conservative estimate for $\sigma_o$, and
0.38\arcsec\ for $\sigma_1$. 

The standard CIAO tool {\em wavdetect} was also used to search for sources 
using an input event file binned by a factor of 4. The 2 sources found with 
{\em wavdetect} were consistent with the sources listed in 
Table~\ref{source_table}. Both sources are outside the 90\% confidence 
4\arcmin\ \rxte error circle for GRO J2058+42. \cxo\ is just 18\arcsec\ outside
the error circle, while I01 is 4.5\arcmin\ from the error circle.

\subsubsection{X-ray Spectral Analysis\label{ss:xray_color_color}}

\begin{figure}[!h]
\includegraphics[angle=270,scale=0.3]{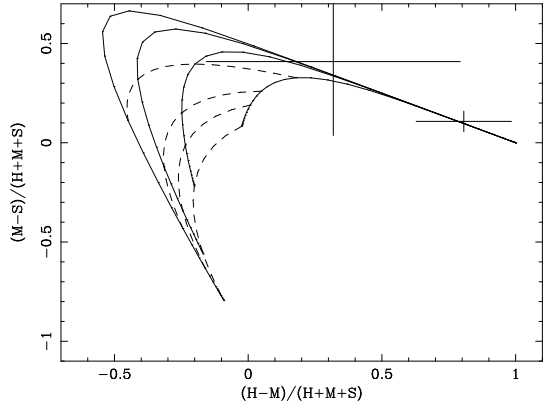}
\figcaption{X-ray ``color-color" diagram for the two sources detected with the
ACIS-I array. 
The energy bands S, M, and H are as follows: S = (0.5 -- 1.0) keV, M = 
(1.0 -- 2.0) keV, and H = (2.0 -- 8.0) keV. The solid lines are contours for 
power-law spectra of constant photon number index ranging from --1 (innermost) to
--4 (outermost) where $N_H$ is varying. The dashed lines are contours of 
constant $N_H$ for a power-law spectrum of varying spectral index. $N_H$ is 0.1,
1, 2, and 5 $\times 10^{21}$ \cmsq\ from the innermost to the outermost 
contour. Thus a source with spectral index --1 and $N_H$ of $10^{20}$ \cmsq\
would be placed on the plot at the intersection of the dashed and solid lines at
approximately (0.0, 0.1). The very hard source to the right is \cxo\ and the 
source farther left is IO1. 
\label{f:xcolors}}
\end{figure}

We show the X-ray ``color-color" diagram in Figure~\ref{f:xcolors}. Since
both sources were detected with only a small number of counts, the
uncertainties are large and it is difficult to draw firm conclusions, although
it is clear that \cxo\ is {\em very hard} and most of the detected flux
appears above the iridium m-edge (circa 2 keV) in the telescope response. 

To extract a spectrum and associated response files for \cxo,
we used a source region consisting of a 5.5\arcsec\ radius circle centered on 
the best-fit \cxo\ position. A 5.5\arcsec\ circle corresponds to the radius that 
encloses 99.9\% of the circular Gaussian that approximately matches the PSF at 
the source location. (See Section~\ref{ss:detect}.) We extracted our spectrum 
from a fits file already filtered to contain only events with energies between 
0.5-8 keV, with the CIAO tool {\em psextract}. The small number of photons
detected did not allow us to use the $\chi^2$ statistic, so we used the Cash
statistic \citep{Cash79} in XSPEC 11.3.1. Since Cash statistics cannot be used 
on background subtracted spectra, we did not subtract a background; however, the
errors introduced from this are likely to be small because we expect about 0.7 
background photons in our source extraction region. We fitted an absorbed 
power-law spectrum to the 64 \chandra counts. The resulting fit parameters were:
$N_{\rm H} = (2.1 \pm 1.0) \times 10^{22}$ cm$^{-2}$ and photon index = 
$1.8 \pm 0.6$. Using the error option on the flux command in XSPEC, we estimated
that the 90\% confidence range for the 2-10 keV unabsorbed flux was $(3-9)
\times 10^{-13}$ \ergcms. The quality of the fit was investigated by generating
10,000 Monte Carlo simulations of the best-fit spectrum. If the model fit the 
data, the fit quality was $\sim 50$\%, meaning that approximately 50\% of the 
simulations had Cash statistics lower than that of our data. For the power-law 
model, the fit quality was 60\%. We also fitted the data with a thermal models, 
a blackbody (phabs*bbodyrad) and a Bremsstrahlung (phabs*bremss), but in
both cases the fit quality was poorer, about 70\% and 80\%, respectively. For 
the blackbody, the emitting region was very small, with a radius of $\sim 0.1$ 
km.

\subsubsection{Timing}
We searched for, but did not detect, pulses from GRO J1958+42 using events 
extracted as discussed in Section~\ref{ss:xray_color_color}. The search was 
conducted in the frequency range of 5.045-5.155 mHz  using the $Z^2_n$ test, 
with the number of Fourier terms, $n$, ranging from 1 to 6. The maximum values 
of $Z^2_n$ found were all consistent with no pulses being present. Monte-Carlo
 simulations were used to determine 95\% confidence upper-limits to the 
pulse fraction of 37\% rms for a simple sinusoidal pulse, and 43\% rms 
for a pulse profile with Fourier power limited to the pulse frequency 
and its first harmonic. For comparison, during the outbursts detected with the
\rxte PCA, the 2-10 keV pulsed fractions were 15-24\% rms.

\subsection{Archival \rosat Observations}

Using a PSF-fitting algorithm, we analyzed archival data from a 2.4 ks \rosat 
High Resolution Imager (HRI) observation, centered on the \rxte position for 
GRO J2058+42, performed on June 23, 1997. Based on the outburst ephemeris of 
\citet{Wilson00}, this observation took place about 8.5 days before the 
predicted GRO J2058+42 outburst peak. No bright transient object was detected. A
total of 5 X-ray sources were detected in the full HRI-FOV and are listed in 
Table~\ref{tab:rosat}. None of these was inside the 4\arcmin\ error circle. The
object nearest to the error circle was object \# 2, and was 0.4\arcmin\ outside
the error circle. This, as well as the other 4 sources, had count rates which 
were below the sensitivity threshold of the \rosat all-sky survey (RASS) done 
between June 1990 to January 1991 (when no source was detected). Objects \#1,2,
and 5 were also found with the standard \rosat analysis and are listed in the 
1RXH catalog. However, the two sources consistent with object \#1 were flagged 
as close to the detector structure and possibly suspect. Sources \#3 \& 4 were 
not found in the catalog, consistent with the lowest significance as reported in
Table~\ref{tab:rosat} and below the significance threshold of the \rosat 1RXH 
catalog. As an additional check, we used the CIAO tool {\em wavdetect} to search
for sources in the \rosat HRI image. Only one source was found, corresponding to
source \#2, with a significance of 4.1 $\sigma$\footnote{The significance was 
found by dividing the net counts by the ``Gehrels error" of the background 
counts, $\sigma = 1 + (BKG\_COUNTS + 0.75)^{1/2}$. See the Detect Manual on
\url{http://asc.harvard.edu/ciao/manuals.html} for details.} consistent with detecting just the 
brightest of the 5 sources. Unfortunately, CIAO does not offer the more 
sensitive PSF-fitting algorithm.
\begin{deluxetable}{ccccccc}
\tabletypesize{\footnotesize}
\tablecaption{Sources Detected with \rosat \label{tab:rosat}}
\tablewidth{0pt}
\tablehead{\colhead{\#} & \colhead{R.A.} & \colhead{Decl.} & 
 \colhead{Total Counts} & \colhead{Rate} & 
  \colhead{Error} & \colhead{Catalog Sources\tablenotemark{a}} \\
 & & & & \colhead{(cts s$^{-1}$)} & \colhead{(cts s$^{-1}$)} & \\}
\startdata  
1 &   314.7504\arcdeg & 41.8786\arcdeg 
 & 11.1 & 0.0049 & 0.0016  
 & 1RXH J205900.2+415242 \\
 \nodata & \nodata & \nodata & \nodata & \nodata & \nodata 
 & 1RXH J205900.0+415242\\
2 &   314.6972\arcdeg & 41.7771\arcdeg & 13.5 & 0.0058 & 0.0016 
 & 1RXH J205847.5+414637 \\
 \nodata & \nodata & \nodata & \nodata & \nodata & \nodata 
 & 1RXH J205847.3+414638 \\
 \nodata & \nodata & \nodata & \nodata & \nodata & \nodata 
 & 1RXH J205847.3+414638 \\
 \nodata & \nodata & \nodata & \nodata & \nodata & \nodata 
 & 1RXH J205847.5+414641 \\   
3 &  314.9085\arcdeg & 41.7064\arcdeg &  6.3 & 0.0027 & 0.0012 & \nodata\\
4 &  314.7629\arcdeg & 41.6315\arcdeg  & 4.3 &   0.0019 &  0.0010 & \nodata \\
5 &  314.8119\arcdeg & 41.6086\arcdeg  & 7.3 &   0.0032 & 0.0013 &
  1RXH J205915.0+413631 \\
\enddata
\tablenotetext{a}{Sources within errors from ``The First \rosat Catalog of Pointed
Observations with the High Resolution Imager (ROSATHRITOTAL / 1RXH)'' available
through \url{http://heasarc.gsfc.nasa.gov} and
\url{http://www.xray.mpe.mpg.de/cgi-bin/rosat/src-browser}.}
\end{deluxetable}
Of the 5 \rosat sources in Table~\ref{tab:rosat}, only \cxo\ (source \#2) was 
detected with \chandrano. Source \#1 fell between the ACIS-I and ACIS-S chips. 
Sources \#3-5 were marginal \rosat detections and unless they were very soft or
transient, should have been detectable by \chandrano. Hence they are likely
spurious.

If the energy spectrum is assumed the same for the \chandra and \rosat
observations of \cxo, the \rosat count rate corresponds to about 10 times
brighter than that observed with \chandrano, indicating that the flux varied significantly
between the two observations. This is similar to GRO J2058+42 (see 
Figure~\ref{fig:asm}) which was actively outbursting in 1997, while the 
outbursts faded below detectability with \rxte by the time of the \chandra 
observation.

\section{\cxo\ Optical/IR Observations and Results}

\subsection{Archival Searches}
We used BROWSE\footnote{see http://heasarc.gsfc.nasa.gov/db-perl/W3Browse/w3browse.pl}
to search for cataloged objects within the 99\% confidence the X-ray error
circles listed in Table~\ref{source_table}. Counterparts were found only for 
\cxo\ in the USNO-B1.0 and 2MASS catalogs. The angular separation between these two objects is
$0.13\arcsec$. The 2MASS object magnitudes were:$J = 11.740 \pm 0.022$, 
$H = 11.282 \pm 0.018$, and $K = 10.930 \pm 0.017$. The 4\arcmin\ \rxte error 
circle contained 605 USNO-B1.0 sources and 636 2MASS sources. The corresponding
densities of sources were used to calculate the expected average number of 
accidental coincidences,  $N_{r99}$, listed in column 5 of Table~\ref{t:counterparts}. The probability of
getting one or more matches by chance is given by the Poisson probability, 
$1-e^{-N_{r99}}$, which for small values of the exponent is approximately 
$N_{r99}$. 
\begin{deluxetable}{llccc}
\tabletypesize{\scriptsize}
   \tablewidth{0pc}
   \tablecaption{Candidate Counterparts to the \chandra X-ray Sources\label{t:counterparts}}
    \tablehead{\colhead{\chandra Name} & \colhead{Catalog} & \colhead{R.A.} & 
    \colhead{Dec.} & \colhead{$N_{r99}$\tablenotemark{a}} \\
    &  & \colhead{(J2000)}   & \colhead{(J2000)} & } 
\startdata 
   I01     &  USNO  & \nodata  &  \nodata & 0.19  \\
   \nodata &  2MASS & \nodata  &  \nodata & 0.20  \\
   \cxo    &  USNO  & 314.698078\arcdeg & 41.777028\arcdeg & 0.021 \\
   \nodata &  2MASS & 314.698057\arcdeg & 41.777000\arcdeg  & 0.022 \\
\nodata\tablenotemark{b} & \nodata & 314.6979\arcdeg & 41.7769\arcdeg & 0.001 \\
\enddata
\tablenotetext{a}{The average number of accidental coincidences expected in
 the 99\% confidence \chandra error circle.}
\tablenotetext{b}{From our optical observations.}
\end{deluxetable}
\subsection{Optical Photometric Observations}

\begin{figure}[!h]
\epsscale{1.0}
\plotone{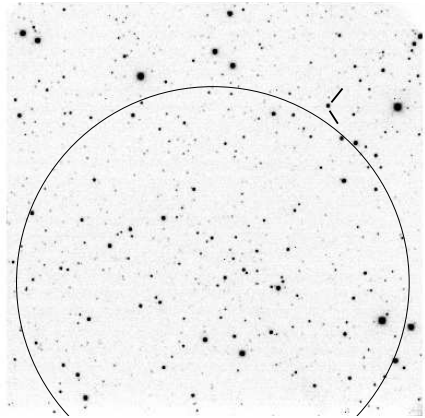}
\caption{$8.4\arcmin \times 8.4\arcmin$ R-band image taken from the
Skinakas Observatory on 2003 June 8. The RXTE error circle, centered on $\alpha
= 314.75\arcdeg$, $\delta = +41.72\arcdeg$ (J2000), and the
proposed optical counterpart, $\alpha = 314.6979\arcdeg$, 
$\delta = +41.7769\arcdeg$ (J2000), are shown. Our 99\% confidence 
1.4\arcsec\ radius \chandra error circle, centered at 
$\alpha = 314.69794\arcdeg$, $\delta = +41.77704\arcdeg$ (J2000) 
falls entirely on the marked optical star in this image. Source I01 and the
bottom edge of the \rxte error circle were outside the field of view.
\label{Rimag}}
\end{figure}

The field around the best-fit \rxte position for GRO J2058+42 was observed 
through the $B$, $V$, $R$, and $I$ filters and a narrow filter centered at 6563 
\AA\ (H$\alpha$ filter) using the 1.3\,m telescope of the Skinakas observatory
on several occasions throughout summer 2003 and 2004 (see
Table~\ref{phot}). The telescope was equipped with a $1024 \times 1024$
SITe CCD chip with a 24 $\mu$m pixel size (corresponding to
$0.5\arcsec$\ on the sky). The telescope was pointed at the \rxte position 
R.A = $314.75\arcdeg$, Decl. = $+41.72\arcdeg$\ and the field of
view was $\sim 8.4\arcmin \times \sim 8.4\arcmin$. Standard stars from the 
\citet{lan92} and \citet{oja96} lists were used for the transformation 
equations.  Reduction of the data was carried out in the standard way using the
IRAF tools for aperture photometry. The results of the photometric analysis are
given in Table~\ref{phot}.
\begin{deluxetable}{lccccc}
\scriptsize
\tablecolumns{6}
\tablecaption{Photometric Measurements of \cxo's Counterpart\label{phot}}
\tablewidth{0pt}
\tablehead{\colhead{Date} & \colhead{MJD} & \colhead{B} & \colhead{V} & 
 \colhead{R} & \colhead{I}}
\startdata
2003 June 7 &52798.50 &16.01$\pm$0.03 &14.91$\pm$0.03 &14.22$\pm$0.03 & \nodata \\
2003 June 8 &52799.44 &16.07$\pm$0.03 &14.94$\pm$0.02 &14.24$\pm$0.02 & \nodata \\
2003 August 24 &52876.37 &16.03$\pm$0.02 &14.90$\pm$0.03 &14.25$\pm$0.03 &13.49$\pm$0.03 \\
2004 July 5 &53192.46 &16.03$\pm$0.02 &14.88$\pm$0.02 &14.16$\pm$0.02 &13.35$\pm$0.02 \\
2004 July 27 &53214.51 &16.07$\pm$0.02 &14.92$\pm$0.03 &14.19$\pm$0.03 &13.41$\pm$0.04 \\
2004 August 24 & 53242.43 & 16.05$\pm$0.02 & 14.95$\pm$0.02 & 14.24$\pm$0.02 &
 13.46$\pm$0.02 \\
\enddata
\end{deluxetable}
Figure~\ref{Rimag} shows a Johnson $R$-band image of the field around GRO
J2058+42 obtained from the Skinakas 1.3-m telescope on 2003 June 8. The
\rxte error circle and the optical counterpart to \cxo\ have
been marked. The second \chandra source, I01 in Table~\ref{source_table}, and
the lower portion of the \rxte error circle were outside the field of view.

\begin{figure}[!h]
\epsscale{1.0}
\plotone{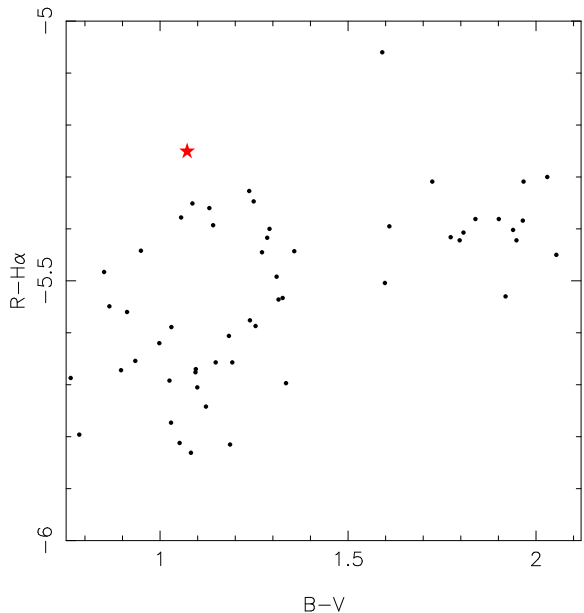}
\caption{Optical color-color diagram of the field around GRO J2058+42. The
data point marked with a star symbol is the \cxo\ optical counterpart.
\label{cd}}
\end{figure}

An optical color-color diagram was generated with the  instrumental
magnitudes of the $B$, $V$, $R$, and H$\alpha$ filters obtained during the
2003 June 8 and 2004 July 5 observations by plotting the ``red color"
$R-H\alpha$ as a function of the ``blue color" $B-V$ (See Figure~\ref{cd}). Be
stars showing emission in H$\alpha$ are expected to occupy the upper left
region of the diagram. Be stars are expected to show low (blue-dominated) 
$B-V$ colors because they are early-type stars (although they normally appear
redder than non-emitting B stars due to the circumstellar disk) and also bright
(less-negative) $R-H\alpha$ colors because they are relatively strong H$\alpha$
emitters.

The color-color diagram taken alone provides two potential candidates to be 
the optical counterpart associated with GRO J2058+52. The two stars lie 
outside the 90\% \rxte error circle, but while the data point marked with a 
star-like symbol ($\alpha = 314.6979\arcdeg$, $\delta = +41.7769\arcdeg$, 
J2000) is 18\arcsec\ outside the error circle and corresponds to the X-ray 
source \cxo\ detected with \chandrano, the other candidate (the one showing the
largest $R-H\alpha$ color values and located at $\alpha = 314.6833\arcdeg$, 
$\delta = +41.7878\arcdeg$, J2000) is about 1.2\arcmin\ away from the error 
circle and was not detected with \chandra or \rosatno. Also, the optical 
spectrum of the other candidate does not resemble that of an early-type star.

\subsection{Optical Spectroscopic Observations}

Optical spectroscopic observations of the \cxo\ counterpart were
obtained from the Skinakas observatory in Crete (Greece) on 2004 June 25,
July 6, and August 26, and from the William Herschel Telescope (WHT) located at
the Roque de Los Muchachos observatory in La Palma (Spain) on 2004 July 4. The 
1.3\,m telescope of the Skinakas Observatory was equipped with a 
1024$\times$1024 Thomson CCD and a 1302 l~mm$^{-1}$ grating, giving a
nominal  dispersion of $\sim$1.3 \AA/pixel. The WHT instrumental set-up
during the service run utilized the ISIS double-arm spectrograph. The blue
arm used the R1200B grating with wavelength coverage 3890 - 4440 \AA,
while the red arm used the R1200R grating with a wavelength coverage of
6180 - 6920 \AA. The spectra were reduced using the STARLINK  {\em Figaro}
package \citep{sho01} and analyzed using the STARLINK {\em Dipso}  package
\citep{how98}.

\begin{figure}[!h]
\epsscale{0.84}
\plotone{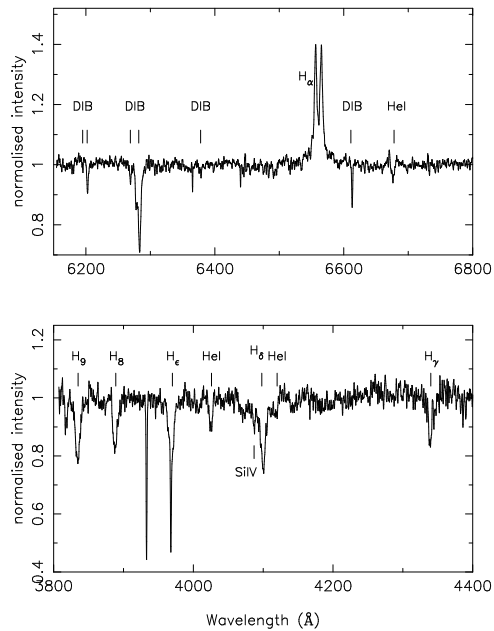}
\caption{The red (top) and blue (lower) spectrum of the counterpart to \cxo\ 
taken on 4 July 2004. A strong, split H$\alpha$ profile is visible with
an equivalent width of $-4.5 \pm 1.0$\AA. The He 6678\AA\ line is also clearly
visible, as are various Diffuse Interstellar Bands. In the blue,
several lines from the Balmer series, the He I lines at 4026\AA\ and 4121\AA\ 
and Si IV 4089\AA\ can be seen in absorption.  \label{fig:red}}
\end{figure}

These spectroscopic observations revealed H$\alpha$ emission with split 
profiles. Equivalent widths (EW) for the H$\alpha$ profiles are listed in
Table~\ref{tab:Ha}. In the red band WHT spectrum, top panel in
Figure~\ref{fig:red}, the He 6678 \AA\ line also shows a double-peaked profile.
The split profiles provide further evidence for a Be  star, since the 
double-peaked shape can be interpreted as coming from the  circumstellar 
envelope. The blue band spectrum, lower panel in Figure~\ref{fig:red} showed 
various other members of the Balmer series in absorption. 
\begin{deluxetable}{lcc}
\tablecolumns{3}
\tablecaption{Equivalent Widths of the H$\alpha$ Line\label{tab:Ha}}
\tablewidth{0pt}
\tablehead{\colhead{Date} & \colhead{MJD} & \colhead{Equivalent Width (\AA)}}
\startdata
2004 June 25 & 53181 & $-4.2\pm0.6$ \\
2004 July 04 & 53190 & $-4.5\pm1.0$\tablenotemark{a} \\
2004 July 06 & 53192 & $-5.0\pm0.9$ \\
2004 August 26 & 53244 & $-4.7\pm0.5$ \\
\enddata
\tablenotetext{a}{Affected by a cosmic ray hit.}
\end{deluxetable}

The blue band spectrum lacks, however, strong He I lines, such as 4009,
4144, and 4387 \AA. Only He I 4026 and 4121 \AA\ (this one was rather
weak) are present. The strength of He I lines reaches a maximum for spectral 
type B2 and falls off on either side (i.e., B1, B3, etc). Thus,
we are likely dealing with a very early B  or late O star. The Si IV 4089 \AA\
line is quite prominent. In fact, the ratio Si  IV 4089 \AA/He I 4121 \AA\
$>$ 1. This could be explained assuming an earlier  spectral type (earlier
than O9) or else with a more evolved companion. The first possibility is
ruled out by the lack of He II (for example, He II 4200 \AA). The lack of
information above 4400 \AA\ makes it difficult to perform a good  spectral
classification, but the data suggest a classification of O9.5-B0IV-V

\section{Discussion}

\subsection{Distance Estimate from Optical/IR data}

In order to estimate the distance it is important to make a
good estimate of the amount of absorption to the source.

The interstellar absorption to the source can be estimated from the
strength of the diffuse interstellar bands \citep[DIB,][]{her75}.
The measurement of the equivalent widths of the DIB is, however,  hampered by 
the low signal-to-noise ratio of the optical continuum. Using the strongest 
lines (those at 6202 \AA\ and 6613 \AA), the estimated color excess is 
$E(B-V)=1.3 \pm 0.1$, but the number of measurements is only 3. If we add the 
5778/80, 6195 and 6269 \AA\ lines then the average value of the reddening is 
$E(B-V)=1.2 \pm 0.2$, where the error is the standard deviation of 7 measurements.
The main uncertainty in the derivation of the equivalent widths comes from the 
difficulty in defining the continuum.

This color excess is consistent with that estimated from the photometric
data. A B0V star has an intrinsic color $(B-V)_0=-0.26$ 
\citep{weg94} and taking the measured photometric color $(B-V)=1.12 \pm 0.04$ we
derive an excess $E(B-V)=1.38 \pm 0.04$. The slightly higher reddening obtained
from the photometric magnitudes may be explained by the contribution of
the circumstellar disk around the Be star. In contrast, the interstellar
lines should be free of such effects.

Using the H$\alpha$ EW $\sim$ 4.5 \AA\ the expected IR colors may be estimated
for a Be star plus circumstellar disk. Using Figure 8 from \citet{Coe94} the 
intrinsic $(J-K)$ is shown to be in the range $-0.03 \pm 0.05$. This gives 
$E(J-K) \sim 0.84 \pm 0.05$ and a corresponding $E(B-V) \sim 1.6 \pm 0.1$. 
Though slightly higher than the value of $1.38 \pm 0.04$ determined from the 
more precise photometry, the agreement is sufficiently good to confirm that both
the optical and IR data are likely to be from the same object.

Taking into account the above three estimates of the reddening we
determine the weighted mean value to be $E(B-V)=1.4 \pm 0.1$.

\begin{figure}[!h]
\epsscale{1.0}
\hspace{-0.75in}
\includegraphics[angle=270,scale=0.4]{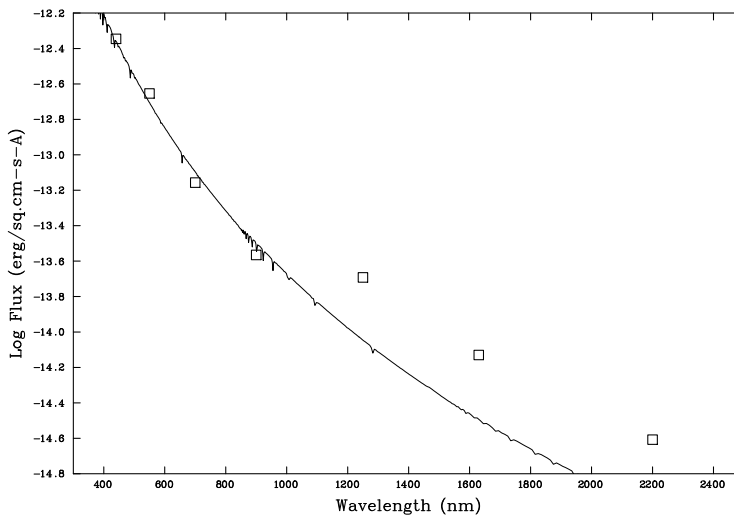}
\figcaption{Fit of Kurucz atmospheric model for a B0/B1 star to the
optical and infrared fluxes de. Squares denote the photometric values (optical 
and infrared) de-reddened by $E(B-V) = 1.4$. The solid line denotes a stellar
atmosphere model for a B0 star. \label{fig:optIR}}
\end{figure}

Figure~\ref{fig:optIR} shows the photometric values (optical and infrared)
de-reddened by E(B-V)=1.4. Superimposed on the these values is a
stellar atmosphere for a B0 star ($T_{\rm eff} = 28,000$K and $\log g = 4.0$)
\citep{Kurucz79}. The model has been normalized to the B band data
since it is assumed that this band will not be significantly affected
by any contribution from the circumstellar disk. It is clear from this
figure that the IR fluxes lie significantly above the model fit,
indicating a strong IR excess. This is to be expected from a Be star,
though the strength of the excess in this case is large - perhaps
partially because the data were not taken
contemporaneously. Nonetheless, the general quality of fit to the
optical fluxes is further confirmation that our estimate for the
reddening is reasonably good.

Taking the standard law $A_V=3.1 E(B-V)$ \citet{Rieke85} and
assuming an average absolute magnitude of $M_V = -4.2 \pm 0.1$ 
\citep{vac96} the distance to GRO J2058+42 is estimated to be $\sim$
$9.0 \pm 1.3$ kpc. The error was obtained by propagating the errors in
$m_V$, $E(B-V)$ and $M_V$. Note that the error in the absolute magnitude
accounts for the uncertainty in the spectral type of the optical companion
(O9.5-B0.5).

The distance estimate from the optical/IR counterpart to \cxo\ is consistent 
with X-ray distance estimate for GRO J2058+42 of 7-16 kpc \citep{Wilson98}, 
based on measurements of spin-up and flux in the giant outburst in 1995. 


\subsection{Comparison with Other Quiescent Be/X-ray Binaries}

From our \chandra observations, \cxo's energy spectrum was approximately
fitted with a power-law with a photon index $\sim 1.8$ and an unabsorbed flux
$F_{1-10\ \rm{keV}} = (3-9) \times 10^{-13}$ \ergcms, or $L_{1-10\ \rm{keV}} = 
(3-9) \times 10^{33}$ \ergss\ for a distance of 9 kpc. \citet{Campana02} 
measured power-law indices of 2.1 and 2.6 and unabsorbed 
$L_{0.5-10\ \rm{keV}} = (1-3) \times 10^{35}$  and $(0.8-2) \times 10^{33}$ \ergss\ for
A0538--66 and 4U 0115+63, respectively, with \BeppoSAXno. No pulsations were 
detected for either source, with a $3 \sigma$ upper limit of 30\% for 4U 
0115+63.  \chandra detected 22 photons from V0332+53 in quiescence corresponding
to $L_{0.5-10\ {\rm keV}} \sim 10^{33}$ \ergcms \citep{Campana02}. No optical 
observations were reported for these objects, so the state of the Be disk was 
unknown. Quiescent observations of A0535+26 with \BeppoSAX \citep{Orlandini04} 
and \rxte \citep{Neg00} are most intriguing. In 1998, A0535+26 was detected at 
$L_{3-20\ \rm{keV}} = \sim 4 \times 10^{33}$ \ergss\ with \rxteno, with a 
power-law index of $\sim 3$. Pulsations were detected during this observation 
with a pulse fraction of $\gtrsim 53$\%. Optical observations of V725 Tau, the 
counterpart to A0535+26, showed H$\alpha$ in absorption from the underlying 
star, indicating that the circumstellar disk was gone \citep{Neg00}. \BeppoSAX 
observations in 2000-2001 also revealed pulsations with a $\sim 50$\% fraction,
a power-law index of $\sim 2$ and an unabsorbed  $L_{2-10\ \rm{keV}} = (1.5-4.4)
\times 10^{33}$ \ergss. However optical observations indicated that H$\alpha$ 
emission, i.e., the circumstellar disk, had returned.  \cxo's luminosity and 
spectral shape are similar to that observed from A0535+26, especially in the 
\BeppoSAX observations where the Be disk had returned. If \cxo\ and GRO
J2058+42 are the same object, it is also a long-period pulsar similar to
A0535+26.

The \rxte ASM light curve showed that regular outbursts of GRO J2058+42 
continued from its discovery in 1995 until at least mid-2002. This long period
of activity from 1995-2002 was similar to EXO 2030+375 \citep{Wilson02}. For 
EXO 2030+375, the extended activity was believed to be the result of the Be 
disk being truncated at a particular resonance radius (4:1), so that an 
outburst was expected at every periastron passage unless the Be disk 
disappeared \citep{Okazaki01}. After mid-2002, GRO J2058+42's outbursts
faded below detectability in the \rxte ASM. If \cxo\ and GRO J2058+42 are the
same object, our optical observations show that although GRO J2058+42's 
outbursts appeared to have ceased by 2004, detection of H$\alpha$ emission 
indicated that the Be disk was still present. Hence GRO J2058+42's behavior, 
while similar to EXO 2030+375's during outbursts, does not fit the EXO 
2030+375 model predictions for outburst cessation. EXO 2030+375's outbursts 
have not ceased to date, so it is not clear whether or not EXO 2030+375 will 
fit the model either.

\section{Conclusions}

Clearly we have discovered a Be/X-ray binary \cxo. Different estimates of 
the number of Be stars in the Galaxy give radically different numbers, but all
give a low chance probability of finding a serendipitous Be star in our \chandra
error circle. Based on evolutionary arguments, \citet{Meurs89} predict 
2000-20,000 Be/X-ray binaries in the galaxy. \citet{Porter03} state that about 
1/3 of known Be stars are in binaries. About 2/3 of these binaries are Be/X-ray
binaries, i.e., they contain a neutron star.  This leads to roughly 
10,000-100,000 Be stars in the Galaxy. If we assume an angular size of 
$360\arcdeg \times 20\arcdeg$\ for the Galactic plane, that implies 0.004-0.004
Be stars per arcmin$^2$. Hence, the chance probability of finding an unrelated 
Be star in our \chandra 99\% confidence error circle is 0.00007-0.0007\%.
Alternatively, the mass of the Galaxy is $1.8 \times 10^{11} M_{\sun}$
\citep{Zombeck90}, meaning about $10^{11}$ stars in the Galaxy. About 1 star in
800 is a B star \citep{Binney98} and about 17\% of B stars are Be stars
\citep{Zorec97}. This gives us a much larger number of Be stars in the Galaxy,
about $2 \times 10^{7}$, leading to about 0.8 Be stars per arcmin$^2$ and a
chance probability of 0.1\% of finding an unrelated Be star in our \chandra error
circle. Neither argument takes into account the fact that not all Be stars show
H$\alpha$ emission at a given time and some percentage of them will not be 
observable due to absorption effects, so the actual probability is even lower. 
In addition, Be stars in Be/X-ray binaries cover a very narrow range in spectral
type \citep[09-B2,][]{Negueruela98} and the spectral type of \cxo's companion is
in that range.

Based on evolutionary arguments \citep{Meurs89}, we should expect to find 
0.004-0.04 Be/X-ray binaries in the \rxte error circle for GRO J2058+42 and 
0.02-0.2 in the ACIS-I field. However, based on our alternative argument, we 
would expect to find about 8 Be/X-ray binaries in the \rxte error circle and 
about 44 in the ACIS-I field. Both estimates assume that all Be/X-ray binaries 
in the galaxy are detectable. To our knowledge to date no previously unknown 
quiescent Be/X-ray binaries have been discovered in X-rays, suggesting that 
either they are less common than predicted or are too faint to observe. Our
observations and observations of other \chandra fields \citep[e.g.,][]{Rogel04},
suggest that the alternative argument overestimates the number of Be stars.
Unfortunately little is known about quiescent Be/X-ray binaries.

Because pulsations were not detected with our \chandra observation, we cannot
definitively say that \cxo\ is GRO J2058+42. However, beyond the positional 
coincidence, other evidence is suggestive that they are the same source. \cxo\ 
was the brightest object observed with \rosat in 1997, when GRO J2058+42 was 
active, and it was roughly 10 times brighter than in the \chandra observations. 
Similarly, \rxte PCA observations showed that GRO J2058+42 was a factor of about
10-100 brighter in outburst in 1998 than upper limits in 2003 December. The 
\rosat observations approximately corresponded in orbital phase to the faintest
\rxte detections. Longer-term observations with the ASM also showed that GRO 
J2058+42 had faded. Further, GRO J2058+42 shows classic Be/X-ray binary behavior
and \cxo\ is associated with a Be star. Lastly, the 7-16 kpc distance to GRO 
J2058+42 is in agreement with the distance of 7.7-10.3 kpc estimated for the 
optical counterpart to \cxo. Additional observations of \cxo\ are needed to 
be certain if it is a new quiescent Be/X-ray binary or if it is GRO J2058+42.

\acknowledgements
This publication makes use of data products from the Two Micron All Sky Survey,
which is a joint project of the University of Massachusetts and the Infrared 
Processing and Analysis Center/California Institute of Technology, funded by 
the National Aeronautics and Space Administration and the National Science 
Foundation. In addition, this research has made use of data obtained from the 
High Energy Astrophysics Science Archive Research Center (HEASARC), provided by NASA's 
Goddard Space Flight Center (GSFC) and quick-look results provided by the 
\rxteno/ASM teams at MIT and at the SOF and GOF at GSFC.


\begin{thebibliography}{}
\bibitem[Apparao(1994)]{Apparao94} 
Apparao, K. M. V. 1994, \ssr, 69, 255
\bibitem[Anders \& Grevesse(1989)]{Anders89}
Anders, E. \& Grevesse, N. 1989, Geochimica et Cosmochimica Acta 53, 197
\bibitem[Arnaud(1996)]{Arnaud96}
Arnaud, K. A. 1996, in ASP Conf. Ser. 101, Astronomical Data Analysis Software
and Systems V, ed. G. Jacoby \& J. Barnes (San Francisco: ASP), 17
\bibitem[Balucinska-Church \& McCammon(1992)]{Bal92}
Balucinska-Church, M. \& McCammon, D. 1992, \apj,  400, 699
\bibitem[Bildsten et al.(1997)]{Bildsten97} 
Bildsten, L., et al. 1997, \apjs, 113, 367 
\bibitem[Binney \& Merrifield (1998)]{Binney98}
Binney, J. \& Merrifield, M. 1998, Galactic Astronomy (Princeton:Princeton
Univ. Press)
\bibitem[Blackburn (1995)]{Blackburn95}
Blackburn, J.K. 1995, in ASP Conf. Ser., Vol. 77, Astronomical Data Analysis and
Software Systems IV, eds. R.A. Shaw, H.E. Payne, \& J.J.E. Hayes (San
Francisco:ASP), 367 
\bibitem[Campana et al.(2002)]{Campana02}
Campana, S. et al. 2002, \apj, 580, 389 
\bibitem[Cash (1979)]{Cash79}
Cash, W. 1979, ApJ, 228, 939 
\bibitem[Castro-Tirado \& Birkle (1996)]{Castro96}
Castro-Tirado, A.J. \& Birkle, K. 1996, \iaucirc, 6516
\bibitem[Coe(2000)]{Coe00}
Coe, M. J., 2000, in IAU Colloq 175, The Be phenomenon in Early-Type Stars,
ed. M.A. Smith \& H.F. Henrichs (ASP Conf. Ser. 214; San Francisco: ASP), 656
\bibitem[Coe et al.(1994)]{Coe94}
Coe, M.J. et al. 1994, \aap, 298, 784
\bibitem[Dickey \& Lockman (1990)]{Dickey90}
Dickey, J.M. \& Lockman, F.J. 1990, \araa, 28, 215
\bibitem[Ghosh \& Lamb (1979)]{Ghosh79}
Ghosh, P. \& Lamb, F.K. 1979, \apj, 234, 296
\bibitem[Grove (1995)]{Grove95}
Grove, J.E. 1995, \iaucirc, 6239
\bibitem[Hanuschik(1996)]{Hanuschik96} 
Hanuschik, R. W. 1996, 
\aap, 308, 170
\bibitem[Henrichs (1983)]{Hen83}
Henrichs, H.F. 1983, in Accretion Driven Stellar X-ray Sources, eds. W.H.G.
Lewin \& E.P.J. van den Heuvel, (New York: Cambridge University Press), 393
\bibitem[Herbig (1975)]{her75}
Herbig, G.H., 1975, \apj, 196, 129
\bibitem[Howarth et al.(1998)]{how98}
Howarth, I.~D., Murray, J., Mills, D., \& Berry, D.~S. 1998, Starlink User Note
50.21
\bibitem[Kurucz (1979)]{Kurucz79}
Kurucz, R.L., 1979, \apjs, 40, 1
\bibitem[Landolt (1992)]{lan92}
Landolt, A. U.,  1992, \aj, 104, 340
\bibitem[Monet et al.(2003)]{Monet03} 
Monet, D.G., Levine, S.E., Casian, B., et al. 2003, \aj, 125, 984
\bibitem[Meurs \& van den Heuvel(1989)]{Meurs89}
Meurs, E.J.A. \& van den Heuvel, E.P.J. 1989, \aap, 226, 88
\bibitem[Negueruela (1998)]{Negueruela98}
Negueruela, I. 1998, A\&A, 338, 505
\bibitem[Negueruela \& Okazaki(2001)]{Negueruela01a} 
Negueruela, I., \& Okazaki, A. T. 2001,
\aap, 369, 108
\bibitem[Negueruela et al.(2001)]{Negueruela01b} 
Negueruela, I., et al. 2001, \aap, 369, 117
\bibitem[Negueruela et al.(2000)]{Neg00}
Negueruela, I., Reig, P., Finger, M.H., Roche, P. 2000, \aap, 356, 1003
\bibitem[Oja(1996)]{oja96}
Oja, T., 1996, BaltA, 5, 103
\bibitem[Okazaki \& Negueruela(2001)]{Okazaki01} 
Okazaki, A.T. \& Negueruela, I. 2001, \aap, 377, 161
\bibitem[Orlandini et al.(2004)]{Orlandini04}
Orlandini, M. et al. 2004, Nuclear Physics B (Proc. Suppl.), 132, 476
\bibitem[Ostriker \& Shu (1995)]{Ost95}
Ostriker, E.C. \& Shu, F.H. 1995, \apj, 447, 813
\bibitem[Porter(1996)]{Porter96}Porter, J. M., 1996, 
\mnras 280, L31
\bibitem[Porter \& Rivinius(2003)]{Porter03}
Porter, J.M. \& Rivinius, T. 2003, \pasp, 115, 1153
\bibitem[Quirrenbach et al. (1997)]{Quirrenbach97}
Quirrenbach, A, et al. 1997, 
\apj, 479, 477
\bibitem[Reig, Kougentakis, \& Papamastorakis (2004)]{Reig04}
Reig, P., Kougentakis, T. \& Papamastorakis, G. 2004, ATel 308
\bibitem[Revnivtsev (2003)]{Rev03}
Revnivtsev, M. 2003, \aap, 410, 865
\bibitem[Rieke \& Lebofsky (1985)]{Rieke85}
Rieke, G.H. \& Lebofsky, M.J. 1985, \apj, 288, 618
\bibitem[Rogel et al.(2004)]{Rogel04}
Rogel, A.B. et al. 2004, astro-ph/0410036
\bibitem[Shortridge et al. (2001)]{sho01}
Shortridge, K., Meyerdierks, H., Currie, M., et~al.
2001, Starlink User Note 86.19
\bibitem[Slettebak(1988)]{Slettebak88}
Slettebak, A. 1988, 
\pasp, 100, 770
\bibitem[Stella et al.(1986)]{Stella86}
Stella, L. et al. 1986, \apj, 308, 669
\bibitem[Swartz et al.(2003)]{Swartz03} 
Swartz, D. A., Ghosh, K.K., McCollough, M.L., Pannuti, T.G.,  Tennant, A.F. \& 
Wu, K. 2003, \apjs, 144, 213
\bibitem[Townsend, Owocki, \& Howarth (2004)]{Townsend04}
Townsend, R. H. D., Owocki, S. P., Howarth, I. D.,  2004, \mnras, 350,189
\bibitem[Vacca, Garmany, \& Shull (1996)]{vac96}
Vacca, W. D., Garmany, C. D., \& Shull, J. M.,  1996, \apj, 460, 914
\bibitem[Valinia \& Marshall (1998)]{Valinia98}
Valinia, A. \& Marshall, F.E. 1998, \apj, 505, 134
\bibitem[Wasserman \& Shapiro (1983)]{Wasserman83}
Wasserman, I. \& Shapiro, S.L. 1983, \apj, 265, 1036
\bibitem[Wegner (1994)]{weg94}
Wegner, W., 1994, \mnras, 270, 229
\bibitem[Weisskopf et al.(2003)]{Weisskopf03}
Weisskopf, M.C. et al. 2003, Experimental Astronomy, 16, 1
\bibitem[Wilson et al.(1995)]{Wilson95}
Wilson, C.A. et al. 1995, \iaucirc, 6238
\bibitem[Wilson et al.(2002)]{Wilson02}
Wilson, C.A. et al. 2002, \apj, 570, 287
\bibitem[Wilson et al.(1998)]{Wilson98}
Wilson, C.A., Finger, M.H., Harmon, B.A., Chakrabarty, D., Strohmayer, T. 1998,
\apj, 499, 820
\bibitem[Wilson, Finger, \& Scott (2000)]{Wilson00}
Wilson, C.A., Finger, M.H., \& Scott, D.M. 2000,  in AIP Conf. Proc. 510,
The Fifth Compton Symposium, ed. M. L. McConnell \& J. M. Ryan
(Melville: AIP), 208
\bibitem[Wilson, Strohmayer, \& Chakrabarty (1996)]{Wilson96}
Wilson, C.A., Strohmayer, T., \& Chakrabarty, D. 1995, \iaucirc, 6514 
\bibitem[Wilson-Hodge(1999)]{WilsonHodge99} Wilson-Hodge, C.A. 1999, PhD 
 Dissertation, University of Alabama in Huntsville 
\bibitem[Yan et al.(1998)]{Yan98}
Yan, M., Sadeghpour, H.R., Dalgarno, A. 1998, \apj, 496, 1044
\bibitem[Zombeck(1990)]{Zombeck90}
Zombeck, M.V. 1990, Handbook of Space Astronomy \& Astrophysics (New York:
Cambridge University Press), 82
\bibitem[Zorec \& Briot(1997)]{Zorec97}
Zorec, J. \& Briot, D. 1997, \aap, 318, 443 
\end{thebibliography}
\end{document}